# An Enhanced Evaluation Model For Vertical Handover Algorithm In Heterogeneous Networks

Mohamed Lahby[1], Leghris Cherkaoui[2] and Abdellah Adib[3]

[1,2,3]Computer Science Department, LIM Lab.
Faculty of Sciences and Technology of Mohammedia,
B.P. 146 Mohammedia, Morocco

**Abstract**
The vertical handover decision is considered an NP-Hard problem. For that reason, a large variety of vertical handoff algorithms (VHA) have been proposed to help the user to select dynamically the best access network in terms of quality of service (QoS).
The objective of this paper is to provide a new approach for evaluating of the vertical handoff algorithms in order to choose the most appropriate algorithm which should be used to select the best access network. Simulation results are presented to evaluate and to test our new evaluation model.

## 1. Introduction

With the evolution of radio access technologies (RAT's) such as wireless technologies (802.11a, 802.11b, 802.15, 802.16, etc.) and cellular networks (GPRS, UMTS, HSDPA, LTE, etc.), the users have the opportunity to utilize a wealth of services across a multitude of these RAT's.

The most important issue in RAT's, is to ensure ubiquitous access for the end users, under the principle "Always Best Connected" (ABC) [1]. To achieve this issue the vertical handoff decision [2] is intended to choose the most suitable network in terms of quality of service (QoS) for mobile users. The vertical handover is the process that transfers call from on base station (BS) or point of attachment (AP) which is base on one of RAT's to another base station which based on different RAT's. This process can be divided into three parts namely: handover initiation, network selection and handover execution.

This work focuses on the network selection step which is the most important key of vertical handover. The network selection problem in heterogeneous wireless networks is complex problem mapped in NP-Hard problem [3], hence it is desirable to use a heuristic algorithm in order to achieve an optimal network selection which can satisfy better tradeoff between network conditions, requirements of applications and users preferences. Three issues dominate the network selection which are a) selecting the appropriate handover metrics, b) identification the most algorithm that exploits these metrics and c) determination the appropriate weighting algorithm that allows to weigh each criterion for each traffic classes.

The handover metrics are the criteria used in the network selection decision to choose the best access network in terms of QoS for the end users. The network selection depends on multiple handover metrics [4] which are:

- From terminal side: battery, velocity, etc.
- From service side: QoS level, security level, etc.
- From network side: provider's profile, current QoS parameters, etc.
- From user side: users preferences, perceived QoS, etc.

In the other hand, several decision algorithms based on multi attributes decision making (MADM) methods have been proposed to deal with the vertical handover algorithm (VHA) problem. The MADM includes many methods such as analytic hierarchy process (AHP), simple additive weighting (SAW), multiplicative exponential weighting (MEW), grey relational analysis (GRA), technique for order preference by similarity to ideal solution (TOPSIS) and the distance to the ideal alternative (DIA). In [5], [6], [7] and [8], the network selection algorithm is based on AHP method and GRA method. The AHP method is used to determine weights for each criterion and GRA method is applied to rank the alternatives. In [9], [10] and [11], the network selection algorithm combines two MADM methods AHP and TOPSIS. The AHP method is used to get weights of the criteria and TOPSIS method is applied to determine the ranking of access network. In [12], the authors present a new MADM method namely DIA to solve the network selection problem. The DIA method selects the alternative that is the shortest euclidean distance to positive ideal alternative (PIA). In [13] the authors propose a novel method based on MADM techniques and mahalanobis distance. The proposed method takes into consideration the correlation between the criteria and aims to choose the optimal network while ensuring no ranking abnormality and reducing the number of handoffs.

In addition, there are several methods used to assign weights for the criteria such as AHP method, fuzzy analytic hierarchy process (FAHP), analytic network process (ANP), fuzzy analytic network process (FANP) and random weighting. Determining the most suitable weights for different criteria for each traffic classes is one of the main problems in the network selection decision. In [14] five weighting algorithms namely AHP, FAHP, ANP, FANP and RW are studied and compared for all four traffic classes namely, conversational, streaming, interactive and background.

Due to the variety of vertical handoff algorithms, variety of the weighting algorithms and also variety of handover metrics, the network selection decision remains a complex issue in the next generation of heterogeneous wireless networks. To achieve this issue, the evaluation of VHA becomes mandatory to reach an optimal network selection algorithm that allows mobile users to choose the best access network with seamless manner.

Some evaluation models for VHA have proposed in the literature. In [15], the authors compare the performance of five VHA, namely SAW, MEW, TOPSIS, GRA, and UA (Abique's Algorithm). Each VHA method used the AHP method to get weights of the criteria and the fuzzy logic is applied to build the evaluation scale and compare different handover metrics. Two traffic classes were considered conversational and streaming. Each traffic class was associated with six handover metrics namely available bandwidth, bit error rate, delay, security and monetary costs.

In [16], the authors compare the performance of three MADM methods namely SAW, MEW and TOPSIS, each method considers five handover metrics namely packet jitter, packet delay, utilization link, packet loss and cost per byte. The performance comparison focuses on three aspects namely the ranking order, the ranking abnormality and the difference on ranking values of all algorithms. In [17], the authors proposed a multi-constraint optimization technique in order to achieve better tradeoff between set of handover metrics such as bit error rate (BER), available bandwidth (ABW) and network traffic (NT). The proposed algorithm is based on the results of performance evaluation parameters namely handoff dropping and call blocking probability. In addition the sensitivity analysis for four traffic classes are presented as follows: the background traffic is sensitive to the ABW, the interactive traffic is sensitive to BER, the conversational traffic and the streaming traffic are sensitive to the NW.

In [18], the authors compare the performance of seven VHA based on MADM methods which are SAW, MEW, TOPSIS, ELECTRE, VIKOR, GRA and WMC (weighted markov chain). The performance evaluation is focused on four parameters of QoS namely packet delay, packet jitter, the available bandwidth and the total bandwidth. Two different applications were considered: voice and data connections. Each traffic application was associated with six attributes: available bandwidth, total bandwidth, packet delay, packet jitter, packet loss and cost per byte.

In [19] and [20] the authors have proposed a new evaluation model for VHA based on multicriteria evaluation and criticality analysis. In one hand the proposed model in [19] allows to evaluate the performance of VHA methods by using the measured values of three parameters namely number of handoffs, handoff delay and computational complexity. In the other hand the proposed evaluation model in [20] is used to evaluate the performance of five MADM methods namely SAW, MEW, TOPSIS, ELECTRE, and VIKOR. For each MADM method we analyze the performance of five parameters namely available bandwidth, delay, jitter, packet loss and cost ber byte.

However, one of the major weaknesses of this model is the lack of a weighting algorithm which can be used to assign a relative weight to each handover metric by considering each traffic classes. To address this issue, we propose to introduce the AHP method in order to find a suitable weight of each performance metric which should be used in specific traffic classes.

This paper is organized as follows. Section II presents the enhancement of the evaluation model. Section III includes the simulations and results. Section IV concludes this paper.

## 2. The enhancement of the evaluation model

In this section, we present an enhancement of the evaluation model proposed in [19] and [20]. The proposed evaluation model combines the multi criteria evaluation and criticality analysis and allows assigning suitable weights for each evaluation parameters by using the AHP method.

The procedure can be categorized in seven steps:

1) Identification of the evaluation parameters: the evaluation parameters represent the indicators that influence the performance of vertical handover algorithm and allow comparing between themes. In this study we use two evaluation parameters namely ranking abnormality and number of handoffs. The ranking abnormality means that the ranking of candidate networks change when low ranking alternatives are removed from the candidate list, which can make the handover vertical algorithm inefficient. The number of handoffs represents the number of network handoffs that the terminal mobile have performed for a given time period.

2) Construct the evaluation matrix: the evaluation matrix is the decision matrix that represents the

evaluation of each vertical handover algorithm Alg$_i$ with respect to the evaluation parameter P$_j$. The evaluation matrix is expressed as:

$$EM = \begin{bmatrix} v_{11} & v_{12} & \cdots & \cdots & v_{1m} \\ v_{21} & v_{22} & \cdots & \cdots & v_{2m} \\ \vdots & \vdots & \vdots & \vdots & \vdots \\ v_{n1} & v_{n2} & \cdots & \cdots & v_{nm} \end{bmatrix} \quad (1)$$

Where v$_{ij}$ is the measured value of the vertical handover algorithm Alg$_i$ with respect to the evaluation parameter P$_j$. The v$_{ij}$ is obtained from simulating vertical handover algorithm by using MATLAB.

3) Construct the normalized evaluation matrix: in order to control the magnitude of evaluation parameters and to prevent that some of the evaluation parameters can dominate others, we calculate the normalized evaluation matrix by Max method normalization. Each element d$_{ij}$ is computed as:

- For benefit attribute, the normalized value of d$_{ij}$ is computed as:

$$d_{ij} = \frac{v_{ij}}{v_j^{max}} \quad (2)$$

- For cost attribute, the normalized value of d$_{ij}$ is computed as:

$$d_{ij} = \frac{v_j^{min}}{v_{ij}} \quad (3)$$

4) Construct the criticality matrix: according to valuation scale defined in table I, we analyze the evaluation matrix obtained in second step. the criticality matrix c$_{ij}$ is computed as:

$$C_{ij} = k \quad (4)$$

Where k is obtained from table 1 according to the value of d$_{ij}$

Table 1: Attribute values for the candidate networks

| Very low k=1 | Low k=3 | Medium k=5 | High k=7 | Very hight k=9 |
|---|---|---|---|---|
| d$_{ij}$ > 80% of the max value | d$_{ij}$ > 60% of the max value | d$_{ij}$ > 40% of the max value | d$_{ij}$ > 20% of the max value | d$_{ij}$ <= 20% of the max value |

5) Construct the weighted criticality matrix: we apply the AHP method to weigh each evaluation parameter, the weighted criticality matrix t$_{ij}$ is computed as:

$$t_{ij} = w_i * c_{ij} \quad where \quad \sum_{i=1}^{m} w_i = 1 \quad (5)$$

6) Calculation of the criticality index: the criticality index of each vertical handover algorithm Alg$_i$ can be calculated as:

$$CI_i = 100 * (\sum_{i=1}^{m} t_{ij})/n \quad where \quad i = 1, ..., n \quad (6)$$

n is the maximum valuation level of all parameters.

## 5. RESULTS AND SIMULATION

### 5.1 The simulation scenario

In this simulation, we consider a heterogeneous environment, which entails six candidate networks, and each network with six parameters. The scenario consists of two 3G cellular networks: UMTS1 and UMTS2, two WLANS: WLAN1 and WLAN2, and two WMANS: WIMAX1 and WIMAX2.

The six attributes associated in this heterogeneous environment. The attributes are: Cost per Byte (CB), Available Bandwidth (AB), Security (S), Packet Delay (D), Packet Jitter (J) and Packet Loss (L).

Table 2: Attribute values for the candidate networks

| Criteria Network | CB (%) | S (%) | AB (mbps) | D (ms) | J (ms) | L (per10$^6$) |
|---|---|---|---|---|---|---|
| UMTS1 | 60 | 70 | 0.1-2 | 25-50 | 5-10 | 20-80 |
| UMTS2 | 80 | 90 | 0.1-2 | 25-50 | 5-10 | 20-80 |
| WLAN1 | 10 | 50 | 1-11 | 100-150 | 10-20 | 20-80 |
| WLAN2 | 5 | 50 | 1-11 | 100-150 | 10-20 | 20-80 |
| WIMAX1 | 50 | 60 | 1-60 | 60-100 | 3-10 | 20-80 |
| WIMAX2 | 40 | 60 | 1-60 | 60-100 | 3-10 | 20-80 |

During the simulation, the measures of every criterion for candidate networks are randomly varied according to the ranges shown in table 2.

### 5.2 Testing of the proposed evaluation model

In order to evaluate and to test the proposed enhanced evaluation model, three vertical handover algorithms based on MADM methods namely TOPSIS, GRA and DIA were presented. We perform four simulations for four traffic classes [21] namely background, conversational, interactive and streaming. In each simulation the three algorithms were simulated in MATLAB to select the suitable vertical handover which should be used in each traffic class.

#### 5.2.1 The simulation 1

In this simulation, the traffic analyzed is background traffic. Table 3 shows the analytical results of three

IJCSI International Journal of Computer Science Issues, Vol. 9, Issue 3, No 2, May 2012
ISSN (Online): 1694-0814
www.IJCSI.org

257algorithms TOPSIS, GRA and DIA. For each vertical handover algorithm we provided the values for average of two performance evaluation ranking abnormality and number of handoffs.

Table 3: Measures for ranking abnormality and number of handoffs

| MADM Algorithms | Ranking abnormality (%) | Number of handoffs (%) |
|---|---|---|
| TOPSIS | 50 | 70 |
| GRA | 20 | 80 |
| DIA | 30 | 60 |

Based on the Table 1, we analyze the evaluation matrix obtained in the Table 3. The results of the analysis between the Table 1 and the Table 3 are shown in the Table 4.

Table 4: The criticality matrix

| MADM Algorithms | Ranking abnormality | Number of handoffs |
|---|---|---|
| TOPSIS | 1 | 1 |
| GRA | 7 | 1 |
| DIA | 5 | 3 |

Before evaluating the performance parameters of each algorithm in order to choose the best vertical handover algorithm between themes, we use the AHP method to calculate the weights of ranking abnormality and number of handoffs. Table 5 presents the associated weights of each performance parameters for background traffic.

Table 5: Comparison matrix and weighting vector of background traffic

| MADM Algorithms | Ranking abnormality | Number of handoffs | Weights |
|---|---|---|---|
| Ranking abnormality | 1 | 1 | 0.5 |
| Number of handoffs | 1 | 1 | 0.5 |

Table 6 shows the scores and the criticality indices of the all algorithms analyzed for background traffic. We notice that GRA and DIA have the highest score, which means that these algorithms have the best performance than TOPSIS.

So for background traffic, it is desirable to use GRA or DIA two MAMD methods in order to select the best access network.

Table 6: Evaluation of VHA for background traffic

| MADM Algorithms | Ranking abnormality | Number of handoffs | Criticality index |
|---|---|---|---|
| TOPSIS | 1 | 1 | 14.29 |
| GRA | 7 | 1 | 57.14 |
| DIA | 5 | 3 | 57.14 |

### 5.2.2 The simulation 2

In this simulation, the traffic analyzed is conversational traffic. Table 7 shows the analytical results of three algorithms TOPSIS, GRA and DIA. For each vertical handover algorithm we provided the values for average of two performance evaluation ranking abnormality and number of handoffs.

Table 7: Measures for ranking abnormality and number of handoffs

| MADM Algorithms | Ranking abnormality (%) | Number of handoffs (%) |
|---|---|---|
| TOPSIS | 36 | 80 |
| GRA | 18 | 60 |
| DIA | 27 | 60 |

Based on the Table 1, we analyze the evaluation matrix obtained in the Table 7. The results of the analysis between the Table 1 and the Table 7 are shown in the Table 8.

Table 8: The criticality matrix

| MADM Algorithms | Ranking abnormality | Number of handoffs |
|---|---|---|
| TOPSIS | 1 | 1 |
| GRA | 5 | 3 |
| DIA | 3 | 3 |

Before evaluating the performance parameters of each algorithm in order to choose the best vertical handover algorithm between themes, we use the AHP method to calculate the weights of ranking abnormality and number of handoffs. Table 9 presents the associated weights of each performance parameters for conversational traffic.

Table 9: Comparison matrix and weighting vector of conversational traffic

| MADM Algorithms | Ranking abnormality | Number of handoffs | Weights |
|---|---|---|---|
| Ranking abnormality | 1 | 1/3 | 0.250 |
| Number of handoffs | 3 | 1 | 0.750 |

Table 10 shows the scores and the criticality indices of the all algorithms analyzed for conversational traffic. We notice that GRA and DIA have the highest score, which means that these algorithms have the best performance than TOPSIS.

So for conversational traffic, it is desirable to use GRA or DIA two MAMD methods in order to select the best access network.

IJCSI
www.IJCSI.org

IJCSI International Journal of Computer Science Issues, Vol. 9, Issue 3, No 2, May 2012
ISSN (Online): 1694-0814
www.IJCSI.org258

Table 10: Evaluation of VHA for conversational traffic

| MADM Algorithms | Ranking abnormality | Number of handoffs | Criticality index |
|---|---|---|---|
| TOPSIS | 1 | 1 | 20.00 |
| GRA | 5 | 3 | 70.00 |
| DIA | 3 | 3 | 60.00 |

5.2.3 The simulation 3

In this simulation, the traffic analyzed is interactive traffic. Table 11 shows the analytical results of three algorithms TOPSIS, GRA and DIA. For each vertical handover algorithm we provided the values for average of two performance evaluation ranking abnormality and number of handoffs.

Table 11: Measures for ranking abnormality and number of handoffs

| MADM Algorithms | Ranking abnormality (%) | Number of handoffs (%) |
|---|---|---|
| TOPSIS | 42 | 70 |
| GRA | 25 | 60 |
| DIA | 33 | 80 |

Based on the Table 1, we analyze the evaluation matrix obtained in the Table 11. The results of the analysis between the Table 1 and the Table 11 are shown in the Table 12.

Table 12: The criticality matrix

| MADM Algorithms | Ranking abnormality | Number of handoffs |
|---|---|---|
| TOPSIS | 1 | 1 |
| GRA | 5 | 3 |
| DIA | 3 | 1 |

Before evaluating the performance parameters of each algorithm in order to choose the best vertical handover algorithm between themes, we use the AHP method to calculate the weights of ranking abnormality and number of handoffs. Table 13 presents the associated weights of each performance parameters for interactive traffic.

Table 13: Comparison matrix and weighting vector of interactive traffic

| MADM Algorithms | Ranking abnormality | Number of handoffs | Weights |
|---|---|---|---|
| Ranking abnormality | 1 | 1/5 | 0.167 |
| Number of handoffs | 5 | 1 | 0.833 |

Table 14 shows the scores and the criticality indices of the all algorithms analyzed for interactive traffic. We notice that GRA and DIA have the highest score, which means that these algorithms have the best performance than TOPSIS.

So for interactive traffic, it is desirable to use GRA or DIA two MAMD methods in order to select the best access network.

Table 14: Evaluation of VHA for interactive traffic

| MADM Algorithms | Ranking abnormality | Number of handoffs | Criticality index |
|---|---|---|---|
| TOPSIS | 1 | 1 | 20 |
| GRA | 5 | 3 | 66.66 |
| DIA | 3 | 1 | 26.66 |

5.5.4 The simulation 4

In this simulation, the traffic analyzed is streaming traffic. Table 15 shows the analytical results of three algorithms TOPSIS, GRA and DIA. For each vertical handover algorithm we provided the values for average of two performance evaluation ranking abnormality and number of handoffs.

Table 15: Measures for ranking abnormality and number of handoffs

| MADM Algorithms | Ranking abnormality (%) | Number of handoffs (%) |
|---|---|---|
| TOPSIS | 60 | 60 |
| GRA | 30 | 70 |
| DIA | 30 | 60 |

Based on the Table 1, we analyze the evaluation matrix obtained in the Table 15. The results of the analysis between the Table 1 and the Table 15 are shown in the Table 16.

Table 16: The criticality matrix

| MADM Algorithms | Ranking abnormality | Number of handoffs |
|---|---|---|
| TOPSIS | 1 | 1 |
| GRA | 5 | 1 |
| DIA | 5 | 1 |

Before evaluating the performance parameters of each algorithm in order to choose the best vertical handover algorithm between themes, we use the AHP method to calculate the weights of ranking abnormality and number of handoffs. Table 17 presents the associated weights of each performance parameters for streaming traffic.

Table 17: Comparison matrix and weighting vector of streaming traffic

| MADM Algorithms | Ranking abnormality | Number of handoffs | Weights |
|---|---|---|---|
| Ranking abnormality | 1 | 1/7 | 0.125 |
| Number of handoffs | 7 | 1 | 0.875 |

IJCSI
www.IJCSI.org

Table 18 shows the scores and the criticality indices of the all algorithms analyzed for streaming traffic. We notice that GRA and DIA have the highest score, which means that these algorithms have the best performance than TOPSIS.

So for background traffic, it is desirable to use GRA or DIA two MAMD methods in order to select the best access network.

Table 18: Evaluation of VHA for streaming traffic

| MADM Algorithms | Ranking abnormality | Number of handoffs | Criticality index |
|---|---|---|---|
| TOPSIS | 1 | 1 | 20 |
| GRA | 5 | 1 | 30 |
| DIA | 5 | 1 | 30 |

## 6. CONCLUSION

In this work, we have proposed an enhanced evaluation model for vertical handover algorithm. The proposed model combines MADM methods and criticality analysis. The AHP method is introduced in this model to find a suitable weight of each performance metric which should be used in specific traffic classes.

The simulation results show that the GRA method has the highest criticality index for all traffic classes namely: background, conversational, interactive and streaming.